\documentclass[12pt]{article} 
 \usepackage{amsmath,amsfonts}
\usepackage{hyperref}
 \usepackage[nosort]{cite} 
\usepackage{graphicx}
\usepackage{psfrag}

\newcommand\be{\begin{equation}}
\newcommand\ee{\end{equation}}
\newcommand{\bea}{\begin{eqnarray}}
\newcommand{\eea}{\end{eqnarray}}

\newcommand{\Tr}{{\rm Tr}}

\newcommand{\cA}{{\mathcal A}}
\newcommand{\cB}{{\mathcal B}}
\newcommand{\cC}{{\mathcal C}}

\newcommand{\ra}{\rightarrow}

\newcommand{\ba}{\begin{eqnarray}}
\newcommand{\ea}{\end{eqnarray}}
\newcommand{\bi}{\begin{itemize}}
\newcommand{\ei}{\end{itemize}}

\newcommand{\p}{\partial}

\newcommand{\Acal}{{\mathcal A}}
\newcommand{\Bcal}{{\mathcal B}}

\newcommand{\N}{{\mathcal N}}

\newcommand{\Ncal}{{\mathcal N}}

\newcommand{\Ocal}{{\mathcal O}}
\newcommand{\Ecal}{{\mathcal E}}

\newcommand{\Dcal}{{\mathcal D}}
\newcommand{\Scal}{{\mathcal S}}
\newcommand{\Lcal}{{\mathcal L}}

\newcommand{\nn}{\nonumber}

\newcommand{\f}{\frac}
\newcommand{\half}{\frac{1}{2}}
\newcommand{\oo}{\frac{1}}

\newcommand{\aslash}[1]{\,\,{\raise.15ex\hbox{/}\mkern-12mu #1}}
\newcommand{\bslash}[1]{\,\,{\raise.15ex\hbox{/}\mkern-9mu #1}}

\renewcommand{\title}[1]{\vbox{\center\LARGE{#1}}\vspace{5mm}}
\renewcommand{\author}[1]{\vbox{\center#1}\vspace{5mm}}

\newcommand{\email}[1]{\vbox{\center\tt#1}\vspace{5mm}}

\begin{document} 
\bibliographystyle{utphys}  

\begin{titlepage}

\rightline{\small{\tt NSF-KITP-08-89}}
\begin{center}

\vskip 2 cm
\centerline{{\Large {\bf Spectral curves, emergent geometry, }}}
\medskip
\centerline{{\Large{\bf  and bubbling solutions for Wilson loops}}}

\vskip 1cm

{
Takuya Okuda }

{\it Kavli Institute for Theoretical Physics, 
University of California

Santa Barbara, CA 93106, USA
}
\email{takuya@kitp.ucsb.edu}

and
\vskip 5mm

{
Diego Trancanelli}

{\it Physics Department, 
University of California

Santa Barbara, CA 93106, USA
}
\email{dtrancan@physics.ucsb.edu}

\vskip 1cm

{\bf Abstract}

\end{center}
\noindent
We study the supersymmetric circular Wilson loops
of $\Ncal=4$ super Yang-Mills in large representations of the gauge group. 
In particular, we obtain 
the spectral curves of the matrix model which captures
the expectation value of the loops. 
These spectral curves are then proven to be precisely the hyperelliptic surfaces
that characterize the bubbling solutions dual to the Wilson loops, thus yielding an example of a geometry emerging from an eigenvalue distribution.
We finally discuss the Wilson loop expectation value from the matrix model and from supergravity.

\end{titlepage}

\tableofcontents 


\section{Introduction}

In gauge theory/gravity correspondences, a saddle point in the gauge theory path integral is expected to represent the space-time geometry in gravity.
Since the saddle point is determined by the dynamics of the gauge theory, the space-time is said to be  {\it emergent}. A notable example of such a phenomenon is the emergence of the sphere of the dual $AdS_5 \times S^5$ geometry from the matrix quantum mechanics governing the  strong coupling dynamics of the constant modes of the  scalars of ${\cal N}=4$ super Yang-Mills compactified on a $S^3$  \cite{Berenstein:2005aa}.\footnote{See \cite{Berenstein:2008me} for a review of subsequent developments and a list of relevant references.}

When an operator is inserted in the gauge theory path integral, the saddle point, as well as the space-time represented by it, gets deformed.
The new space-time develops bubbles of new cycles carrying fluxes.
Such bubbling geometries were originally found for
half-BPS local operators in $\Ncal=4$ super Yang-Mills theory in the context of the AdS/CFT 
correspondence \cite{Berenstein:2004kk,Lin:2004nb}.
They were later generalized to include Wilson loops \cite{Yamaguchi:2006te,Lunin:2006xr,D'Hoker:2007fq} and surface operators \cite{Gomis:2007fi} of the $\N=4$ theory, while
bubbling in topological string theory was found and studied in 
\cite{Gomis:2007kz,Gomis:2006mv,Halmagyi:2007rw}.

The current work revisits the bubbling geometries for circular supersymmetric Wilson loops in $\N=4$ super Yang-Mills. These geometries were constructed in a complete form in reference \cite{D'Hoker:2007fq}.
The ten-dimensional space-time is a warped product
\bea
ds^2 = f_1 ^2 ds^2 _{AdS_2} + f_2 ^2 ds^2 _{S^2} + f_4^2 ds^2 _{S^4}
+ ds^2 _\Sigma \label{warped-metric}
\eea
 of 
$AdS_2\times S^2\times S^4$ and a half-plane $\Sigma$.
The radii $f_1,f_2,f_4$ and all other supergravity fields are functions on $\Sigma$ given in terms of two holomorphic functions, 
 ${\cal A}$ and $\Bcal$.
In fact, $\Sigma$ is naturally identified with
the lower half-plane in one sheet of a hyperelliptic surface,
also denoted by $\Sigma$, and $\Acal$ and $\Bcal$ are
constructed geometrically.
Thus the data ($\Sigma, \,
{\cal A,\,  B}$) completely characterize the bubbling solution.

In this paper, we demonstrate that the deformed saddle points in gauge theory represent the bubbling geometries by making use of a matrix model.
It was conjectured in \cite{Erickson:2000af,Drukker:2000rr} 
that the Wilson loop expectation value is captured by the Gaussian matrix model with a loop operator insertion.
The conjecture was recently proved in reference \cite{Pestun:2007rz}, where 
it was also shown that the matrix is the constant mode of a scalar field.\footnote{
In \cite{Bonelli:2008rv} it was argued that
the matrix model arises as a mirror 
of the topological A-model for the $AdS_5\times S^5$ superstring 
\cite{Berkovits:2007rj}.
}
We show that the saddle point configuration of the matrix eigenvalues back-reacts to the operator insertion and the hyperelliptic surface $\Sigma$ arises as the spectral curve in a generalized sense that we explain in detail.\footnote{
It was originally argued by Yamaguchi \cite{Yamaguchi:2006te} that the eigenvalue distribution of the matrix model characterizes the bubbling geometry.
}
We also find an interpretation of $\Acal$ and $\Bcal$
in the matrix model.

\begin{figure}[tb]
\begin{center}
\psfrag{n1}{$n_1$}
\psfrag{K1}{$K_1$}
\psfrag{n2}{$n_2$}
\psfrag{K2}{$K_2$}
\psfrag{Kg-1}{$K_{g-1}$}
\psfrag{ng}{$n_g$}
\psfrag{Kg}{$K_g$}
\psfrag{ng+1}{$n_{g+1}$}
\includegraphics[width=80mm]{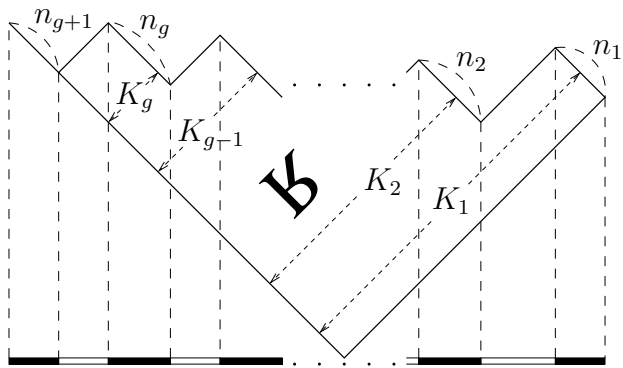} 
\caption{The Young tableau $R$ is shown rotated and inverted.
It consists of $g$ blocks, the $I$-th one of them having $n_I$ rows of length $K_I$.
We set $K_{g+1}\equiv 0$ and $n_{g+1}\equiv N-\sum_{I=1}^g n_I$.
}
\label{parametrization}
\end{center}
\end{figure}

Concretely,
the circular supersymmetric Wilson loop is defined as
\ba
W_R=\Tr_R\, P \exp\oint ( i A+\theta^i \phi^i ds).
\ea
Here $A$ is the  gauge field and $\phi^i$ are the six real scalars.
The integral is along a circle in $\mathbb{R}^4$, $ \theta^i$ is a constant 
unit vector
in $\mathbb{R}^6$, and $s$ is the parameter of the circle such that $||dx/ds||=1$.
The trace is taken in an irreducible representation $R$ of $U(N)$ or $SU(N)$.
Such $R$ is specified by a Young tableau, which is also denoted by the same symbol $R$.
 The dual bubbling geometry has small curvature when  $R$ 
 has long edges and it is characterized by a genus $g$ hyperelliptic surface $\Sigma$, where $g$ is the number of blocks in $R$ (see fig. \ref{parametrization}).
The Wilson loop expectation value is given by the matrix integral
\ba
\langle W_R \rangle_{YM}=\oo{\cal Z}\int [dM] e^{-\f{2N}{\lambda} \Tr M^2} \Tr_R e^{M}.
\ea
The $N\times N$ matrix $M$ is hermitian  and ${\cal Z}$ is the partition function.
For representations $R$ that give rise to smooth bubbling geometries, we solve the matrix model in the limit
where $N$ is infinite and the 't Hooft coupling
$\lambda\equiv g^2_{YM}N$ is large.
As it turns out, $\Acal$ and $\Bcal$ are simply related to 
the resolvent $\omega(z)$
and the spectral parameter $z$ of the matrix model:
\ba
\Acal\propto \omega(z)-2z\, ,
\qquad 
\Bcal\propto z+{\rm const.}\, \label{AB-relation}
\ea
We also show that the resolvent is given by the indefinite integral of a meromorphic 1-form $\alpha$ on the same hyperelliptic surface $\Sigma$.
The surface $\Sigma$ is given by the equation
\ba
y^2=H_{2g+2}(z)\, ,
\ea
and the 1-form $\alpha$ by
\ba
\alpha=\p \omega=\left(2-2\f{a_{g+1}(z)}{\sqrt{H_{2g+2}(z)}}\right)dz\, .
\ea
The polynomials $H(z)$ and $a(z)$ have degrees $2g+2$ and $g+1$ respectively.
We find from the matrix model analysis a set of constraints that determine the coefficients of $a(z)$ and $H(z)$ uniquely.
These constraints are identical to the ones that arise in the bubbling geometry.
The surface $\Sigma$ is the spectral curve of the matrix model in the sense that the eigenvalue distribution is determined by $\Sigma$ and $\alpha$.

Given our large $N$ solution of the matrix model, the Wilson loop expectation value can be easily computed.
A natural question is whether it can also be reproduced in supergravity, by evaluating the 
on-shell action in the bubbling geometry background.
We include in this paper some relevant calculations that will be useful for this purpose.
In particular, we show that the on-shell supergravity Lagrangian is always a total derivative.
This would imply that the on-shell action  splits 
into two contributions, one coming from the new cycles 
of the bubbling geometry and the other  given 
as a surface integral on the conformal boundary.
It is the former contribution that we manage to compute exactly
within an ansatz we make.
This work does not address the latter contribution,
which seems to require a holographic renormalization technology beyond the one currently available.
Indeed, because the new cycles mix non-trivially  the $AdS_5$ and $S^5$
directions,
usual counter-terms in five-dimensional supergravity
cannot be used, at least in a straightforward way.

It is however possible to use the identification of
the matrix model and supergravity data to compare
the correlators of the Wilson loop with local operators, 
namely chiral primaries and the energy-momentum tensor. 
This is reported in a companion paper \cite{correlators}.

We structure the paper as follows.
In Section \ref{sec-curve}, we study the matrix model for Wilson loops dual to bubbling geometries.  We solve the model, obtain its spectral curve, and show that it is the hyperelliptic surface that characterizes the bubbling geometry dual to the Wilson loop.  
Section \ref{sec-action} then focuses on the Wilson loop expectation value.
Using our solution, we compute the Wilson loop expectation value for representations that correspond to smooth bubbling geometries.
  This reproduces the result of \cite{Okuda:2007kh} in a certain limit.
 We next show that the on-shell supergravity Lagrangian is a total derivative and compute the contributions from the new cycles that appear in the bubbling geometry.
We then conclude the paper by discussing the outlook in Section \ref{sec-conclusion}.
The appendices contain details  used in the text.


\section{Spectral curves and bubbling geometries}\label{sec-curve}

The expectation value of  a circular
Wilson loop in $\Ncal=4$ super Yang-Mills is
captured by a Gaussian matrix model \cite{Erickson:2000af,Drukker:2000rr,Pestun:2007rz}.
This was originally proposed for half-BPS loops in the fundamental representation (which are dual to fundamental strings in the bulk), but the conjecture has later been extended and checked to hold also for circular loops in arbitrary representations $R$ of the gauge group \cite{Drukker:2005kx,Yamaguchi:2006tq, Okuyama:2006jc,Hartnoll:2006is,Giombi:2006de} and for some loops preserving reduced amounts of supersymmetry \cite{Drukker:2006ga,Drukker:2006zk,Drukker:2007dw,Drukker:2007yx,Drukker:2007qr}.
The precise statement is that the Wilson loop expectation value for the $U(N)$ theory 
is given by
\ba
\left\langle W_{ R} \right\rangle_{U(N)}
=
\oo {\cal Z} \int [dM] \exp\left({-\f{2N}\lambda \Tr M^2}\right)
\mbox{Tr}_{ R} e^M. \label{mat-wil-U} 
\ea
Here $M$ is an hermitian matrix and the partition function ${\cal Z}$ of the matrix model is
defined as the integral without the insertion $\mbox{Tr}_{ R} e^M$.
We use the standard hermitian measure $[dM]$.
In the absence of operator insertions, the eigenvalues
are distributed in the large $N$ limit according to the 
Wigner semi-circle law.\footnote{
Pedagogical references on general matrix models include
\cite{Ginsparg:1993is,Francesco:1993nw}.}

To make better contact with the supergravity solution, it turns out to be more convenient to follow the procedure delineated in \cite{Halmagyi:2007rw} and decompose $M$ in $g+1$ sub-blocks $M^{(I)}$ of size $n_I \times n_I$.
The  expectation value  of the loop is then given by 
several Gaussian matrix integrals correlated by interactions between the sub-blocks:
\ba
\langle W_{R}\rangle_{U(N)}
=
\oo {\cal Z}
\int \prod_{I=1}^{g+1} [dM^{(I)}] e^{-\f{2N}\lambda \sum_I\Tr( M^{(I)})^2}
e^{K_I\Tr M^{(I)}}
\prod_{I<J}\det\f{(M^{(I)}\otimes 1-1\otimes M^{(J)})^2}
{1-e^{-M^{(I)}}\otimes e^{M^{(J)}}}.
\nn\\
\label{YMmodel}
\ea
The eigenvalues of $M^{(I)}$ for fixed $I$ are distributed
along some interval $[e_{2I},e_{2I-1}]$.
In the following, we drop  the exponential interactions
by replacing $(1-e^{-M^{(I)}}\otimes e^{M^{(J)}})$ with  $1$.
This is a consistent approximation in the limit
\ba
\lambda\gg 1\,,\qquad g_{YM}^2 n_I=\Ocal(\lambda)\,,\qquad g_{YM}^2 (K_I-K_{I+1})=
\Ocal( \lambda^{1/2})\,,
\ea
because $e_{2I-1}-e_{2I}=\Ocal(\sqrt{g_{YM}^2 n_I})$
and $e_{2I}-e_{2I+1}=\Ocal(g_{YM}^2(K_{I}-K_{I+1}))$
as one can see from the saddle point equations below.

Going to the eigenvalue basis, the matrix model in (\ref{YMmodel}) becomes (here $i=1,\ldots,n_I$ labels the eigenvalues of the $I$-th sub-block)
\ba
\langle
W_R
\rangle_{U(N)}
&\propto&
\int \prod_{I,i}dm^{(I)}_i\exp\Bigg[
-\f{2N}\lambda \left(m^{(I)}_i\right)^2
+K_I m^{(I)}_i
\Bigg]
\prod_{(I,i)<(J,j)}
\left[
m^{(I)}_i-m^{(J)}_j
\right]^2.\cr && \label{model-to-solve}
\ea
We have introduced a linear ordering in the set
of all the eigenvalues so that the last product
is taken over distinct pairs of eigenvalues.
The saddle point equations are
\ba
-\f{4N}\lambda m^{(I)}_i+K_I +2\sum_{(J,j)\neq (I,i)}
\oo{m^{(I)}_i-m^{(J)}_j}=0.
\label{spe}
\ea
By defining the resolvent
\ba
\omega(z)\equiv g_{YM}^2 \sum_{(I,i)}\oo{z-m^{(I)}_i},\label{omega}
\ea
the eqs. (\ref{spe}) can be written, for $x\in [e_{2I},e_{2I-1}]$, as
\ba
-4 x+g_{YM}^2 K_I+\omega_+(x)+\omega_-(x)=0\, , 
\label{saddle-point}
\ea
where $\omega_\pm (x)\equiv \omega(x\pm i \epsilon)$.


\subsection{A hyperelliptic surface as the spectral curve}
\label{sec-hypercurve}

By differentiating eq. (\ref{saddle-point}), one can see that $\omega'_{\pm}=4-\omega'_{\mp}$, so that the combination
\bea
\omega'(4-\omega')
\eea
is invariant when crossing the cut. 
Let us now consider the behavior of this expression close to a branch point, 
say $e_1$.
The eigenvalues are expected to produce square root branch cuts.
Since $\omega(z)$ satisfies eq. (\ref{saddle-point}), locally it is given by
\bea
\omega\sim 2z-\frac{1}{2}g_{YM}^2 K_1+c \sqrt{z-e_1},
\eea
where $c$ is some constant.  Then
\bea
\omega'(4-\omega')\sim \left(2+\frac{c}{2\sqrt{z-e_1}}\right)\left(2-\frac{c}{2\sqrt{z-e_1}}\right)= 4-\frac{c^2}{4(z-e_1)}.
\eea
The same behavior is found for every 
branch point $e_i$:
\bea
\omega'(4-\omega')\sim \frac{C_i}{z-e_i} ~~\hbox{ as }~~ z\ra e_i\,,
\qquad C_i={\rm const.},
\eea
so the combination
\bea
\omega'(4-\omega')-\sum_{i=1}^{2g+2}\frac{C_i}{z-e_i}
\eea
is regular everywhere on the complex plane.
The first term behaves as 
$\Ocal(1/z^2)$ for large $z$ by the definition of $\omega$.
Thus
the combination must vanish everywhere and, in addition,
the second term has to be of the form
\bea
-\sum_{i=1}^{2g+2}\frac{C_i}{z-e_i}=- \frac{f_{2g}(z)}{H_{2g+2}(z)}\, , 
\eea
with $f_{2g}(z)$ a polynomial of degree $2g$
and
\bea
H_{2g+2}(z)\equiv \prod_{i=1}^{2g+2}(z-e_i)\,.
\eea
The solution to the quadratic equation
\bea
\omega'(4-\omega')=
\frac{f_{2g}(z)}{H_{2g+2}(z)}\label{omega'-eq}
\eea
is then
\bea
\omega'=2-\sqrt{4-\frac{f_{2g}(z)}{H_{2g+2}(z)}}\equiv 
 2-2\frac{a_{g+1}(z)}{\sqrt{H_{2g+2}(z)}}. \label{omega'-eq2}
\eea
Here we have selected the negative sign in front of the square root to guarantee the correct behavior for $z\to\infty$. 
In introducing the monic  polynomial $a_{g+1}(z)=z^{g+1}+\ldots$, we noted 
that $H_{2g+2}-f_{2g}/4$ has to be a perfect square, 
so that the only singularities of $\omega'$ are the branch points $e_i$.

We can geometrically interpret eq. (\ref{omega'-eq2}) by saying that
the resolvent $\omega(z)$ is the indefinite integral
\ba
\omega(z)=\int^z_\infty \alpha \label{omega-alpha}
\ea
of a meromorphic 1-form
\ba
\alpha= \left(2-2\frac{a_{g+1}(z)}{\sqrt{H_{2g+2}(z)}}\right)\, dz
\label{one-form}
\ea
 on the hyperelliptic curve defined by 
\bea
y^2=H_{2g+2}(z)\, . \label{spec-curve}
\eea
The only singularity of the 1-form $\alpha$ is the double pole at $z=\infty$ on the second sheet.


\subsection{Parameters and constraints}

The parameters in the definition of the spectral curve 
and the one-form are 
the $3g+3$ coefficients of the two monic polynomials $a_{g+1}\equiv a$ 
and $H_{2g+2}\equiv H$.
Let us study the constraints that determine these parameters.

\begin{figure}[tb]
\begin{center}
 \psfrag{A1}{$A_1$}
 \psfrag{A2}{$A_2$}
 \psfrag{A3}{$A_3$}
 \psfrag{B1}{$B_1$}
 \psfrag{B2}{$B_2$}
 \psfrag{e1}{$e_1$}
 \psfrag{e2}{$e_2$}
 \psfrag{e3}{$e_3$}
 \psfrag{e4}{$e_4$}
 \psfrag{e5}{$e_5$}
 \psfrag{e6}{$e_6$}
\includegraphics[width=100mm]{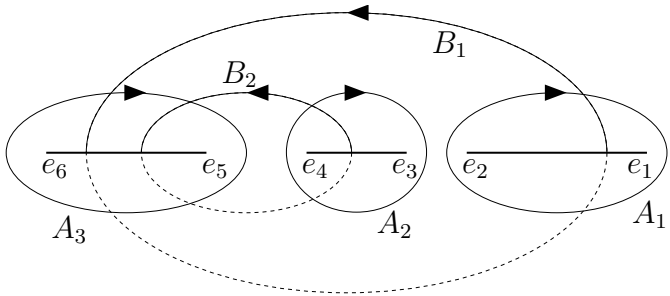} 
\caption{
The $A$- and $B$-cycles of the hyperelliptic surface $\Sigma$
of genus $g=2$.
}
\label{ABcycles}
\end{center}
\end{figure}

The constraints are most concisely expressed in terms of period integrals,
so let us introduce the $A$- and $B$-cycles of the hyperelliptic 
surface in the standard way (see fig. \ref{ABcycles}):
the cycle
$A_I$ (with $I=1,\ldots, g+1$) circles the $I$-th cut $[e_{2I},e_{2I-1}]$
clockwise.
Only the first $g$ of the $A$-cycles are independent, since $A_{g+1}=-A_1-\ldots-A_g$.
The cycle $B_I$ (with $I=1, \ldots, g$) goes through the $I$-th and 
the $(g+1)$-th cuts and has intersection numbers  $\#(A_I\cap B_J)=\delta_{IJ}$ for $J=1,2,\ldots, g$. 

\begin{enumerate}

\item
The first $g+1$ constraints come from 
the requirement that the resolvent $\omega(z)$ should be
single-valued on the physical sheet.
Since it is obtained by integrating the one-form $\alpha$, we need that
\bea
\oint_{A_I}\alpha=0, \qquad I=1,\ldots, g+1.  \label{constraint-1}
\eea
These $g+1$ constraints are all independent: even though
$\sum_{I=1}^{g+1} A_I$ is a trivial cycle in homology,
the condition $\int_{\sum A_I}\alpha=0$ applied to (\ref{one-form})
is non-trivial and ensures that no logarithmic term appears in the expansion
of $\omega$ around $z=\infty$.

\item

According to the saddle point equations (\ref{saddle-point}), the 
value of $\omega$ along  the cycle $B_I$ goes from $\omega$ to $4z-\omega$
 in passing through the $(g+1)$-th cut from the first to the second sheet (recall that $K_{g+1}=0$), and then from $4z-\omega$ to $\omega+g_{YM}^2 K_I$ when coming back to the first sheet across the $I$-th cut.
In terms of the one-form $\alpha$, we get $g$ conditions
\bea
\oint_{B_I}\alpha=g_{YM}^2 K_I\,,\qquad I=1,\ldots,g.
\label{constraint-2}
\eea

\item Since the $I$-th cut contains $n_I$ eigenvalues,
the definition (\ref{omega}) implies the following $g+1$ conditions
\bea
\oint_{A_I}\omega \, dz=-2\pi i g_{YM}^2 n_I\,,\qquad I=1,\ldots, g+1.
\label{constraint-3}
\eea
The integral should be performed on the first sheet.

\item The $3g+2$ conditions above determine $a_{g+1}(z)$ and $H_{2g+2}(z)$
up to a shift of $z$.
The last condition that fixes this ambiguity is
\ba
\omega(e_{2g+2})=2e_{2g+2}\, ,
\label{constraint-4}
\ea
which follows form (\ref{saddle-point}) recalling that $K_{g+1}=0$.

\end{enumerate}

We check now that $\omega(z)$ given by (\ref{omega-alpha})
together with the constraints (\ref{constraint-1})-(\ref{constraint-4})
automatically satisfies the saddle point equations (\ref{saddle-point}).
For this we need to evaluate $\omega$ just above and below each branch cut
$[e_{2I},e_{2I-1}]$.
Since we know the value of $\omega$ at $z=e_{2g+2}$,
we only need to integrate $\alpha$ from $e_{2g+2}$
to $e_{2I}$ along an arbitrary path on the first sheet,
and then from $e_{2I}$ to $x\pm i\epsilon$ with $x\in [e_{2I},e_{2I-1}]$ along the cut.
The key points are that
\ba
4\int_{e_{2g+2}}^{e_{2I}}\f{a(z)}{\sqrt{H(z)}}dz=g_{YM}^2 K_I\,,
\ea
as follows from the condition (\ref{constraint-2}), and that
\ba
\sqrt{H(x+i\epsilon})=-\sqrt{H(x-i\epsilon})
\ea
on the cut.
For $x\in [e_{2I},e_{2I-1}]$ we have
\bea
\omega_+(x)+\omega_-(x)
&=&2\omega(e_{2g+2})+2\int_{e_{2g+2}}^{e_{2I}}
\left(
2-2\frac{a}{\sqrt{H}}
\right) dz
\nn\\
&&~~+\int_{[e_{2I},x]+i\epsilon} 
\left(2-2\frac{a(x')}{\sqrt{H(x')}}\right)dx'
+\int_{[e_{2I},x]-i\epsilon} \left(2-2\frac{a(x')}{\sqrt{H(x')}}
\right) dx'
\nn\\
 &=&4e_{2g+2}+4(e_{2I}-e_{2g+2})-g_{YM}^2 K_I+4(x-e_{2I})
\nn\\
&=&4x-g_{YM}^2 K_I\,,
\eea
so we see that the saddle point equations (\ref{saddle-point}) are satisfied.
Thus at this point we have found the exact solution of the
matrix model (\ref{model-to-solve}) in the large $N$ limit.


\subsection{Comparison}

What remains to be shown is that the spectral curve (\ref{spec-curve})
is the hyperelliptic surface that appears as part of the
bubbling solution for a Wilson loop \cite{D'Hoker:2007fq}.

The bubbling geometry is a warped product of
$AdS_2\times S^2\times S^4$ and a half-plane, as we have mentioned in the introduction.
This half-plane is taken to be the
lower half-plane in one sheet of the hyperelliptic surface
given by
\bea
s^2=\prod_{i=1}^{2g+1}(u-\tilde e_i). \label{surface-su}
\eea
The branch points of the surface are at $u=\tilde e_i$ (with $i=1,\ldots, 2g+1$)
and $u=\tilde e_0\equiv \tilde e_{2g+2}\equiv \infty$.
(Notation changed from \cite{D'Hoker:2007fq}: $e_i^{\rm there}=\tilde e_i^{\rm here}$.)
The constant $u_0$ and the branch points $\tilde e_i$ are
all real and ordered as follows:
\ba
\tilde e_{2g+1}<\tilde e_{2g}<\ldots<\tilde e_1<u_0.
\ea
Though the RHS of (\ref{surface-su}) is a polynomial
of degree $2g+1$ instead of $2g+2$,
the equation can be transformed to the form (\ref{spec-curve})
by a M\"obius transformation on $u$.

All the supergravity fields are expressed in terms of 
two holomorphic functions $\Acal$ and $\Bcal$ on $\Sigma$ given by
\ba
\p \Acal&=&-i\f{P(u)du}{(u-u_0)^2 s(u)},
\\
\Bcal&=&-i\oo{u-u_0}.
\ea
The polynomial $P(u)$ has real coefficients and is of degree $g+1$.
The real part of $\Acal$ must vanish on $[\tilde e_{2I+1},\tilde e_{2I}]$
to ensure regularity of the solution,
so there are constraints
\ba
\int_{[\tilde e_{2I},\tilde e_{2I-1}]-i\epsilon}
\p \Acal&=&0,\qquad I=1,\ldots, g+1.
 \label{bub-con-1}
\ea
The branch cuts $[\tilde e_{2I-1},\tilde e_{2I-2}]$ represent three-cycles
of topology $S^3$ that arise from the geometric transition
of D5-branes, so they carry RR three-form fluxes.
Since each column in the Young tableau $R$ represents a D5-brane
\cite{Yamaguchi:2006tq,Gomis:2006sb}, the flux carried by the
$I$-th cycle is proportional to $K_I-K_{I+1}$, the number of
columns in the $I$-th block:
\ba
8\pi i\int_{[\tilde e_{2I-1},\tilde e_{2I-2}]} \p \Acal+c.c.
=\int_{S^3}F_{(3)}^{\rm RR}=4\pi^2 (K_I-K_{I+1})\alpha' \label{bub-con-2}
\ea
for $I=1,\ldots, g$.
Similarly, the segment $[\tilde e_{2I},\tilde e_{2I-1}]$
represents a five-cycle of topology $S^5$ that arises
from the geometric transition of $n_I$ D3-branes \cite{Drukker:2005kx,Gomis:2006sb} and
carries RR five-form flux.
As we show in Appendix \ref{sec-5form}
\ba
8\pi^2 i\int_{[\tilde e_{2I},\tilde e_{2I-1}]-i\epsilon}
(\Acal\p\Bcal-\Bcal\p\Acal)+c.c.=\int_{S^5}F_{(5)}
=4\pi^4\alpha'^2 n_I 
 \label{bub-con-3}
\ea
for $I=1,\ldots, g+1$.

Shifting the imaginary part of 
$\Acal$ does not affect the physical fields.
It is natural to fix this ambiguity by requiring that
\ba
\lim_{u\ra \infty}\Acal=0.
 \label{bub-con-4}
\ea

The constraints (\ref{bub-con-1})-(\ref{bub-con-4})
are equivalent to (\ref{constraint-1})-(\ref{constraint-4}), respectively,
if we make the identification
\ba
\omega-2z=i\f{8}{\alpha'}g_s\Acal,
\qquad \Bcal=i\f{\alpha'}4 (z-e_{2g+2}).\label{omega-A}
\ea
Equivalently, we have
\ba
\Acal=i\f{\alpha'}{4g_s}\int_{e_{2g+2}}^z \f{a(z')}{\sqrt{ H(z')}}dz',
\qquad u-u_0=\f{4}{\alpha'}\oo{e_{2g+2}-z}.
\ea
Note that $g_{YM}^2=4\pi g_s$.
Thus we have showed that the spectral curve of the matrix
model is precisely the hyperelliptic surface
that characterizes the bubbling geometry.


\subsection{\texorpdfstring{$SU(N)$}{SU(N)} gauge group}

So far we have focused on the $U(N)$ gauge group case.
It is easy to describe what changes for a $SU(N)$ theory.
First, the Wilson loop expectation value of the gauge theory
is related to the matrix model by a simple modification of
(\ref{mat-wil-U}): 
 \ba
\left\langle W_{ R} \right\rangle_{SU(N)}
=
\oo {\cal Z} \int [dM] \exp\left({-\f{2N}\lambda \Tr M^2}\right)
\mbox{Tr}_{ R} e^{M'}, \label{mat-wil-SU}
\ea
where
\ba
M'=M-\oo N (\Tr M) 1_{N\times N}
\ea
is the traceless part of $M$.
Since
\ba
\Tr_R e^{M'}= e^{-\f{|R|}N\Tr M}\Tr_R e^M,
\ea
the saddle point equation (\ref{saddle-point})
for the $I$-th cut
becomes
\ba
-4 x+g_{YM}^2 (K_I-|R|/N)+\omega_+(x)+\omega_-(x)=0.
\label{saddle-point-SU}
\ea
Therefore the resolvents of the $U(N)$ and $SU(N)$ theories
are simply related by a shift of the argument:
\ba
\omega_{SU(N)}(z)=\omega_{U(N)}(z+|R|/4N). 
\ea
Equivalently, the eigenvalue distribution
is simply shifted by a constant so that
the average position of the eigenvalues is the origin.
The relations between $\omega, \, z$ and $\Acal, \, \Bcal$ become
\ba
\omega-2(z+|R|/4N)=i\f{8}{\alpha'}g_s\Acal,
\qquad \Bcal=i\f{\alpha'}4 (z-e_{2g+2}),
\label{SU-ident}
\ea
where $e_{2g+2}\equiv e_{2g+2}^{U(N)}+|R|/4N$ is the last branch point in the $SU(N)$ case.


\section{Wilson loop expectation value}
\label{sec-action}

Given our identification of the matrix model and supergravity data,
it is natural to compare various physical quantities
computed on both sides.
A companion paper \cite{correlators} 
studies the correlators of Wilson loops with local operators, such as chiral primaries and the energy-momentum tensor, finding agreement between gauge theory and supergravity analysis.
Another natural quantity to compare is the Wilson loop expectation value,
which we study in this section.
On the Yang-Mills side, we compute it using our large $N$
solution of the matrix model.
We also discuss the supergravity computation though we do not 
complete it in this paper.\footnote{The computation of the expectation value of a loop dual to D3 and D5 branes \cite{Gomis:2006sb,Gomis:2006im} has been performed in \cite{Drukker:2005kx,Yamaguchi:2006tq,Okuyama:2006jc,Hartnoll:2006is}, both using the matrix model and the DBI action.}
First we prove that the on-shell supergravity Lagrangian
is always a total derivative.
Then we show that  the action contains contributions from
the new cycles of the bubbling geometry and also
from the boundary of space-time.
We compute the first kind of contributions.
Issues with the second type are
discussed in Section \ref{sec-conclusion}.


\subsection{Wilson loop expectation value from  the matrix model} 

To the leading order in the saddle point approximation,
the normalized Wilson loop expectation value is given
by 
\ba
\langle W_R\rangle=e^{-({\cal S}_{\rm mat}-{\cal S}_0)}, \label{S-S}
\ea
where ${\cal S}_{\rm mat}$ and ${\cal S}_0$ are the on-shell actions
of the Gaussian matrix model with and without Wilson loop insertion.
We now proceed with computing these actions.

Again, we begin with the case of a $U(N)$ gauge group.
The on-shell value of the matrix model action is given by
\ba
-{\cal S}_{\rm mat}
&=&
\sum_{I,i}\left[
-\f{2N}\lambda 
 \left(m^{(I)}_i\right)^2
+K_I m^{(I)}_i
\right]
+
\sum_{(I,i)<(J,j)}\log\left[
m^{(I)}_i-m^{(J)}_j
\right]^2
\nn\\
&=&
N\sum_I \int_{[e_{2I},e_{2I-1}]}
 dx\rho(x)
\left[
-\f{2N}\lambda 
x^2
+K_I x
\right]
+N^2
\int_\mathbb{R} dx dy \rho(x)\rho(y)\log|x-y|, \nn\\
\label{double-integral}
\ea
where
the eigenvalue density
\ba
\rho(x)=\oo N\sum_{I,i}\delta(x-m^{(I)}_i)
\ea
is related to the resolvent by
\ba
\rho(x)=\f{i}{2\pi \lambda} (\omega_+(x)-\omega_-(x))\, , \qquad 
\omega(z)=\lambda \int_\mathbb{R} dx \f{\rho(x)}{z-x}.
\ea
In the limit in which  the cuts are well separated,
the last term in (\ref{double-integral}) can be dropped, and by using the eigenvalue
density $\rho(x)$ given by
\ba
\sum_I\f{n_I}{N}\delta\left(x-g_{YM}^2 K_I/4\right),
\ea
we easily reproduce the results of \cite{Okuda:2007kh}.

The expression (\ref{double-integral}) may
be enough for comparison with supergravity
although we do not see
how the double
integral can appear in gravity.
We now rewrite (\ref{double-integral})
in a form that involves no double integral.
First, let us use 
the density and a principal value
integral to re-express (\ref{spe}):
\ba
-4 x +g_{YM}^2 K_I
+ 2 \lambda {\rm P}\int_\mathbb{R} dy \rho(y) \oo{x-y}
=0~\hbox{ for } ~x\in [e_{2I-1},e_{2I}].
\ea
This equation can be integrated to yield
\ba
-2x^2+g_{YM}^2K_I x
+ 2 \lambda \int_\mathbb{R} dy \rho(y) \log|x-y|
=2g_{YM}^2 c_I~~\hbox{ for } ~~x\in [e_{2I-1},e_{2I}],\label{cI}
\ea
where $c_I$ is an integration constant.
The on-shell  matrix action is then
\ba
-{\cal S}_{\rm mat}
&=&N\sum_{I=1}^{g+1} \int_{[e_{2I},e_{2I-1}]} dx\rho(x)
\left[
-\f{N}\lambda 
x^2
+\half K_I x+c_I
\right]. \label{S-cI}
\ea
One expression for the Wilson loop expectation value that does not involve a double integral
or $c_I$ is obtained
by using (\ref{cI}) with $x=e_{2I-1}$ and $x=e_{2I}$:
\ba
&&\log \langle W_R\rangle_{U(N)}
\nn\\
&=&
N\sum_{I=1}^{g+1} \int_{[e_{2I},e_{2I-1}]}
dx\rho(x)
\Bigg[
-\f{N}\lambda 
x^2
+\half K_I x
-\f{N}{2\lambda} (e_{2I-1}^2 +e_{2I}^2)
+
\f{K_I (e_{2I-1}+e_{2I})}4
\nn\\
&&\hskip 1cm+\half \sum_J n_J
\log(e_{2J-1}-x)(x-e_{2J})\Bigg]
-\log\sqrt\lambda+3/4+\log 2\, ,
\ea
where we used 
\ba
{\cal S}_0= N^2\left(-\log\sqrt\lambda+3/4+\log 2\right)
\ea
that follows from the density $\rho_0(x)=(1/\pi\lambda)\sqrt{\lambda-x^2}$ 
for  Wigner's distribution.

For the $SU(N)$ theory, we simply replace $K_I$ by
$K_I-|R|/N$:
\ba
\log \langle W_R\rangle_{SU(N)}
&=&
N\sum_{I=1}^{g+1} \int_{[e_{2I},e_{2I-1}]}
\hspace{-10mm} dx\rho(x)
\Bigg[
-\f{N}\lambda 
x^2
+\half K_I x
-\f{N}{2\lambda} (e_{2I-1}^2 +e_{2I}^2)
\cr &&+
\f{(K_I-|R|/N) (e_{2I-1}+e_{2I})}4
+\half \sum_J n_J
\log(e_{2J-1}-x)(x-e_{2J})\Bigg]
\cr && -\log\sqrt\lambda+3/4+\log 2. \label{SU-action}
\ea
In this formula $\rho(x)$ and $e_i$
are the density and the branch points in the $SU(N)$
case, and we have used the fact that the
average eigenvalue vanishes to remove a shift of $K_I$
in the second term inside the bracket.


\subsection{Wilson loop expectation value from supergravity}

Let us now turn to supergravity.
The solution in \cite{D'Hoker:2007fq} is for an infinite straight line along the Lorentzian time, whereas the matrix model model computation is appropriate for a circle in Euclidean signature. This is not a problem, since both the straight line and the circle preserve the same isometry $SO(2,1)\times SO(3)\times SO(5)$ (albeit  realized differently in the two cases). We can then extend the solution of \cite{D'Hoker:2007fq} to the circular case via a Wick rotation,  considering a fibration with the Euclidean factor $\mathbb{H}_2$, rather than $AdS_2$. This difference will not play any significant role in our analysis, so that we shall consider for simplicity the Lorentzian signature. 
The Wilson loop expectation value is then given by $\langle W_R\rangle=\exp(- \Scal_E)$
after the Wick rotation that identifies $-\Scal_E$ with $i\Scal_L$, where $\Scal_E$ and $\Scal_L$ are
the Euclidean and Lorentzian on-shell actions.


\subsubsection{The on-shell Lagrangian is a total derivative}

We begin our discussion of the supergravity action
by showing that the on-shell Lagrangian density
always has to be a total derivative, if it is a homogeneous
function of the fields of non-zero degree.
It seems well-known that the supergravity Lagrangian
is a total derivative if the equations of motion are satisfied,
though we do not know a reference that makes the general statement
explicitly.

The argument is simple.
Suppose the Lagrangian $\Lcal(\phi)$
depends on the  fields $\phi^i$ and their derivatives.
There can be second or higher derivatives.
When we take the variation of $\Lcal$ with respect
to arbitrary changes $\delta\phi^i$,
in general we get terms that contain derivatives
of $\delta \phi^i$.
By definition, the equations of motion $\Ecal_i(\phi)=0$ are obtained
by rewriting $\delta \Lcal$ as
\ba
\delta \Lcal= \sum_i \delta\phi^i {\cal E}_i(\phi)
+ \sum_i \Dcal_i(\delta\phi^i;\phi),
\ea
where 
$\Dcal_i$ is the total derivative term that is linear in $\delta \phi^i$.
If the Lagrangian is homogeneous, there are (usually integers)
numbers 
$n_\Lcal$ and $n_i$ such that
\ba
\Lcal(\Omega^{n_i} \phi^i)=\Omega^{n_\Lcal}\Lcal(\phi^i)
\ea
for any constant $\Omega$.
We call $n_i$ the dimensions of the fields.
By choosing $\Omega=1+\epsilon$ so that $\delta \phi^i=\epsilon \,n_i\phi^i$,
we find that
\ba
\epsilon\, n_\Lcal \Lcal(\phi)
=\sum_i \epsilon \,n_i \phi^i \Ecal_i(\phi)+\sum_i \Dcal_i(\epsilon\, n_i\phi^i;\phi).
\ea
If the equations of motion are satisfied, the Lagrangian is a total derivative:
\ba
\Lcal(\phi)=\sum_i\f{n_i}{n_\Lcal}\Dcal_i(\phi^i;\phi).
\ea

We now apply the above consideration to the type IIB supergravity action\footnote{Self-duality of the five-form, $F_{(5)}=\star F_{(5)}$, does not follow from this action, 
but has to be imposed by hand. 
One can consider other actions 
where self-duality is implemented with an auxiliary field.
In the case \cite{Dall'Agata:1997ju} we looked at, the on-shell value does not seem to change. }
\bea
2\kappa^2\, {\cal S}&=&\int d^{10}x \, \sqrt{- g}\left( R-\frac{1}{2}\frac{\partial_M\tau\partial^M \bar \tau}{\left(\mbox{Im}\, \tau\right)^2}
\right)\cr 
&&+\int\left(
-\frac{1}{2} M_{ab} H^{a}_{(3)}\wedge  \star H^{b}_{(3)} 
-4 F_{(5)}\wedge \star F_{(5)} 
 -\epsilon_{ab} C_{(4)}\wedge  H_{(3)} ^{a} \wedge H_{(3)} ^b\right).
\label{sugra-action}
\eea
The action is written essentially in the convention of \cite{D'Hoker:2006uu}
and contains various combinations of the fields:
\bea
&&
\tau=C_{(0)}+i e^{-\varphi}
\, , \qquad 
(M_{ab})={\rm diag}(e^{-\varphi},e^{\varphi})
\, , 
\nn\\
&&
F_{(5)}=d C_{(4)}+\oo 8 \epsilon_{ab}B_{(2)}^a\wedge d B_{(2)}^b,
\eea
where $H_{(3)}^a=dB_{(2)}^a$ and $a={\rm NS, \,RR}$.
First note that the action is homogeneous of degree 8
if we assign dimension 2 to the metric $g_{MN}$
and $p$ to all $p$-form fields (scalars are zero-forms).
So our argument applies.
Since the scalars have vanishing dimensions, we can ignore
their variations.
Then under arbitrary variations of the fields except the scalars,
the action changes as
\ba
2\kappa^2 \delta {\cal S}
&=&
\int d^{10}x \sqrt{- g}\nabla^M (\nabla^N\delta g_{MN}-
g^{PQ}\nabla_M \delta g_{PQ})
\nn\\
&&
+
\int d\bigg(
-M_{ab}\delta B_{(2)} ^a \wedge \star H_{(3)} ^b
 -2\epsilon_{ab}C_{(4)}\wedge  \delta B_{(2)} ^a\wedge H_{(3)} ^b
\nn\\
&&~~~~~~~~~~~
-8 \delta C_{(4)}  \wedge \star F_{(5)}  +\epsilon_{ab}\delta B^a_{(2)}
\wedge  B_{(2)} ^b \wedge\star F_{(5)} 
\bigg)
\ea
up to terms that vanish on-shell.
By setting 
\ba
\delta g_{MN}=2\epsilon g_{MN},~~~\delta B_{(2)}^a=2 \epsilon  B_{(2)}^a,
~~~\delta C_{(4)}=4\epsilon   C_{(4)}
\ea
and using $\delta \Scal=8\epsilon \Scal$,
we conclude that
the on-shell action is given by
\ba
2\kappa^2 {\cal S}
\hspace{-.5mm}
&=&
\hspace{-.5mm}
\int d\left(
\hspace{-1mm}
-\oo 4 M_{ab} B_{(2)} ^a \wedge \star H_{(3)} ^b
 -\half \epsilon_{ab}C_{(4)}\wedge   B_{(2)} ^a\wedge H_{(3)} ^b
-4 C_{(4)}  \wedge \star F_{(5)}
\right).
\ea
We thus see that we only need the two- and four-form fields
to compute this part of the action.
Note that so far we have not committed to any particular
solution.


\subsubsection{Contributions from new cycles}

In the solution of \cite{D'Hoker:2007fq},
the NS two-form is along the $AdS_2$ directions
 while the RR two-form along the $S^2$ directions. 
The RR four-form has two components,
one in the $AdS_2\times S^2$ and the other in the $S^4$ directions.
One has then
\bea
B_{(2)}^{\rm NS}=b_1 \hat e ^{01}\, ,\quad
B_{(2)}^{\rm RR}=b_2 \hat e ^{23}\, ,\quad 
C_{(4)}=-j_1 \hat e^{0123}+j_2 \hat e^{4567},
\label{ansatz-forms}
\eea
where $\hat e^{01},~\hat e^{23}$, and $\hat e^{4567}$
are the volume forms of
$AdS_2, \,S^2$, and $S^4$, respectively, all normalized to unit radius.
Note that $b_1,\,b_2,j_1$, and $j_2$ are real functions on $\Sigma$.
Recall now that the $S^2$ and $S^4$ radii
vanish on segments of the real axis of $\Sigma$.
Thus $\hat e^{23}$ and $\hat e^{4567}$ are not
globally defined forms in the ten-dimensional space-time,
while $\hat e^{01}$ is.
This implies that the Chern-Simons term in (\ref{sugra-action})
is not globally defined.
We can make it globally defined by adding further
total derivative terms
\ba
2\kappa^2 \Scal_1=
\int
d\left(2C_{(4)}\wedge  B_{(2)} ^{\rm NS} \wedge H_{(3)} ^{\rm RR}
-\oo{16}B_{(2)} ^{\rm NS}\wedge B_{(2)} ^{\rm NS}\wedge d\left(B_{(2)} ^{\rm RR}\wedge B_{(2)} ^{\rm RR}\right)\right)
\ea
so that the new Chern-Simons term in $2\kappa^2(\Scal+\Scal_1)
\equiv 2\kappa^2 \Scal_{\rm bulk}$ is
\ba
\int  2
F_{(5)} \wedge B_{(2)} ^{\rm NS} \wedge   H_{(3)} ^{\rm RR}.
\ea
The on-shell action is then given by
\ba
2\kappa^2 \Scal_{\rm bulk}
&=&
\int d\left(
-\oo 4 M_{ab} B_{(2)}^a \wedge\star H_{(3)}^b
-B_{(2)}^{\rm NS}
 \wedge B_{(2)}^{\rm RR}\wedge dC_{(4)}
\right), \label{action-der1}
\ea
where we took into account (\ref{ansatz-forms}).

Since some of the forms in (\ref{action-der1})
are not globally defined,  we need caution
in applying the Poincar\'e lemma.
Some terms  in (\ref{action-der1})
are contributions of the non-trivial  cycles in the bubbling geometry,
while the rest are from the boundary of space-time.
We focus on the former contributions.
The latter should be combined with counter-terms we do not discuss
in the present work.

With our ansatz, the Hodge duals of the three-forms are given by
\ba
\star F_{(3)}^{\rm NS}&=&\f{f_2^2 f_4^4}{f_1^2} 
\star db_1\wedge \hat e^{234567},
\\
\star F_{(3)}^{\rm RR}&=&\f{f_1^2 f_4^4}{f_2^2}   \star db_2\wedge \hat e^{014567}.
\ea
We have by the Poincar\'e lemma
\ba
2\kappa^2 \Scal_{\rm bulk}
&=& V\int_{\partial\Sigma}\left(
-\oo 4e^{-\varphi}
\f{f_2^2 f_4^4}{f_1^2} b_1 \star db_1
-\oo 4e^{\varphi}\f{f_1^2 f_4^4}{f_2^2}  b_2 \star  db_2
- b_1 b_2 d j_2 \right).
\label{stokes}
\ea
By $\p\Sigma$ we denote the real axis as well as a large
semi-circle on the lower half-plane.
We cannot meaningfully separate contributions from 
the two components of $\p\Sigma$
because adding an exact form in the integrand of (\ref{stokes})
mixes them.
In (\ref{stokes}) we have made the important assumption that the volume
of $AdS_2$ is regularized in a way independent of the position
on $\p\Sigma$.
We have denoted the volume of $AdS_2\times S^2\times S^4$ by $V$.
In a more complete calculation of the on-shell action,
this assumption may need to be modified.

In Appendix \ref{sec-asymptotic}, we study how various quantities in (\ref{stokes})
behave in the asymptotic region $z\ra\infty$.
If we choose the coordinate 
to be the spectral parameter $z$ in the $SU(N)$ case,
both $b_1$ and $b_2$ vanish as $z\ra \infty$
while $j_2$ remains finite.
Thus the contribution from the semi-circle in this parametrization
vanishes.

On the real axis, the first term in (\ref{stokes}) never contributes 
because it contains positive powers of radii of the two spheres 
and always vanishes.
The remaining two terms nicely combine to give
\ba
V
\int_{-\infty}^\infty 
b_2\left(\oo 4e^{\varphi}\f{f_1^2 f_4^4}{f_2^2} \star  db_2
+ b_1 d j_2 \right). \label{real-contr1}
\ea
The sign change from (\ref{stokes}) is due to the natural
direction for integration.
We observe that $f_4$ vanishes on regions of the real axis
where $S^4$ shrinks to zero size.
In fact, $j_2$ is constant there because otherwise
$F_5$ that contains $dj_2\wedge\hat e^{4567}$ would be ill-defined.
On the other hand, $b_2$ is constant on regions where $S^2$ shrinks
for a similar reason and, as we recall in Appendix \ref{sec-geom},
$b_2=-4 {\rm Im}\,\Acal$.
Since $\Acal(e_{2g+2})=0$ by (\ref{bub-con-4}), the flux condition
(\ref{bub-con-2}) determines these constants to be
\ba
b_2=2\pi \alpha'K_I~~\hbox{ on }~~ [e_{2I},e_{2I-1}].
\ea
Thus we can write (\ref{real-contr1}) as
\ba
V
\sum_{I=1}^{g+1}
\f{\pi}2 \alpha'K_I
\int_{[e_{2I},e_{2I-1}]}
\left(e^{\varphi}\f{f_1^2 f_4^4}{f_2^2} \star  db_2
+ 4b_1 d j_2 \right). \label{real-contr2}
\ea
 
The physical meaning of the integrand in
(\ref{real-contr2}) can be understood as follows.
The equation of motion for $B_{(2)}^{\rm RR}$ can be written as
\ba
d H_{(7)}^{\rm RR}=0,
\ea
where
\ba
H_{(7)}^{\rm RR}\equiv e^{\varphi}\star H_{(3)}^{\rm RR}
+4 B_{(2)}^{\rm NS}\wedge
F_{(5)}
-\half B_{(2)}^{\rm NS}\wedge B_{(2)}^{\rm NS}\wedge H_{(3)}^{\rm RR}.
\ea
It is easy to see that the integrand in
(\ref{real-contr2}) is proportional to the component of $H_{(7)}^{\rm RR}$
along the $AdS_2\times S^4$ direction.
The seven-form 
is to be regarded as the field strength
of the six-form potential $H_{(7)}^{\rm RR}=dC_{(6)}^{\rm RR}$.
By the symmetry of $AdS_2\times S^2\times S^4$,  we can write
\ba
C_{(6)}^{\rm RR}=b_4 \hat e^{014567},
\ea
where $\hat e^{014567}$ is the volume form of
unit $AdS_2\times S^4$.
Then by definition 
\ba
e^{\varphi} \f{f_1^2 f_4^4}{f_2^2} \star db_2
+4b_1 dj_2
=db_4. \label{b4-eqn}
\ea
Thus the integrand in (\ref{real-contr2}) is $db_4$.

One can express the LHS of (\ref{b4-eqn}) in terms of $\Acal$ and $z$
using the known expressions for fields summarized in
Appendix \ref{sec-geom}.
It is in fact possible to integrate the equation:
\bea
\frac{1}{  \alpha'^2}b_4&=&\frac{2(z-\bar z)(\Acal+\bar \Acal)^2-(z^2-\bar z^2)(\Acal+\bar \Acal)(\p_z \Acal+\p_{\bar z} \bar \Acal)}
{2(\p_z \Acal-\p_{\bar z}\bar \Acal)}\cr
&&\hskip 1cm+ \frac{3}{2}(z^2-\bar z^2)(\Acal-\bar \Acal)
-6\int dz\, z \Acal-6\int d\bar z\, \bar z\bar \Acal,\label{b4-answer}
\eea
where the last two terms involve indefinite integrals.
One can check that (\ref{b4-eqn}) is satisfied by this solution.
On the real axis where $z=\bar z$, $b_4$ reduces to
\ba
b_4&=&
-6 \alpha'^2\int dz\, z\,\Acal+c.c.\, .
\ea
Thus
\ba
b_4(e_{2I-1})-b_4(e_{2I})&=&-6\pi^2\alpha'^3 N\int_{[e_{2I},e_{2I-1}]} 
dx  \rho(x) x\, . 
\ea
By collecting everything together,
(\ref{stokes}) becomes
\ba
2\kappa^2 \Scal_{\rm bulk}/V
&=&
-\f{3}2 \pi^3\alpha'^4 N
\sum_{I=1}^{g+1}
K_I
\int_{[e_{2I},e_{2I-1}]} 
dx  \rho(x) x
\, .\label{action-bub}
\ea
This is the contribution from the bulk, in particular from the
cycles that have grown in the bubbling geometry.
This is not the complete story, since the volume $V$
should be regularized 
and counter-terms on the boundary should be added.
We see indeed that (\ref{action-bub}) seems to account only  for
special terms in the matrix model action in  (\ref{SU-action}).


\section{Conclusion}
\label{sec-conclusion}

The main achievement of this paper is the large $N$ solution
of the matrix model that governs circular BPS Wilson loops
at strong coupling.
We determined the eigenvalue distribution for an arbitrary
representation in terms of geometric data on the spectral curve.
The spectral curve was then identified with the
hyperelliptic surface $\Sigma$ that was found in \cite{D'Hoker:2007fq}
to characterize the bubbling geometry for the Wilson loop.

The identification of the hyperelliptic surface $\Sigma$ as
a spectral curve is important for two reasons.
First, one can view this as an example of emergent geometry.
The matrix model is a reduction of the four-dimensional
gauge theory \cite{Pestun:2007rz} and the geometry
emerges out of the dynamics of the eigenvalues.

Second, the identification provides the precise dictionary
between field theory and gravity.
Indeed it serves as the basis for the matching of
physical quantities computed on both sides.
A successful example of matching is reported in \cite{correlators},
where the correlators of the Wilson loop with chiral primaries and
the energy-momentum tensor are computed.

It should also be possible to match the computations
of the Wilson loop expectation value.
Given our solution of the matrix model, we were able to compute
the Wilson loop expectation value quite easily.
On the other hand, the computation of the expectation value
in supergravity is unfinished.
Such computation should involve two non-trivial tasks.
One is to properly take into account the new cycles that
appear in the bubbling geometry.
In the present work, we developed techniques to perform this task.
The other task is to regulate the infinite volume
of the ten-dimensional space-time and to add proper counter-terms.
Usual five-dimensional counter-terms do not suffice, because
the bubbling geometry mixes the $AdS_5$ and $S^5$ directions
in a topologically non-trivial way.\footnote{A similar problem, related to the difficulties of formulating higher-dimensional counter-terms \cite{Taylor:2001fe}, was already encountered by one of the present authors in the context of bubbling geometries \cite{Giombi:2005zq}.}
Construction of the counter-terms is a worthwhile open problem
that has applications to other observables such as
surface operators \cite{Drukker:2008wr}.


\subsection*{Acknowledgments}

We are happy to thank David Berenstein, Eric D'Hoker, Sean Hartnoll, Gary Horowitz, 
Don Marolf, David Mateos, Joe Polchinski, and especially Kostas Skenderis for fruitful discussions. We are also grateful to Jaume Gomis and Shunji Matsuura for collaboration on a related project. This research is supported by the NSF grants 
PHY05-51164 and PHY04-56556, and by the Department of Energy under Contract DE-FG02-91ER40618.


\appendix

 
\section{Details on the bubbling geometry}
\label{sec-geom}

The solution to the BPS equations can be expressed in terms of 
two holomorphic functions $\Acal$ and $\Bcal$
on the lower half-plane.
Let us define four harmonic functions $h_1,\, \tilde h_1,\, h_2,$ and $ \tilde h_2$
by
\ba
\Acal\equiv\half(h_1-i\tilde h_1)\, , \qquad \Bcal\equiv\half(h_2-i\tilde h_2).
\ea
In fact, all the physical fields except the form fields
can be written in terms of $h_1$ and $h_2$ alone.
The field strengths of the form fields are also
given in terms of $h_1$ and $h_2$.
The dual harmonic functions $\tilde h_1$ and $\tilde h_2$
only appear in the potentials \cite{D'Hoker:2007fq}.

It is useful to define the following shorthand notations
\bea
&& V  =  \p _w h_1 \p _{\bar w} h_2 - \p _{\bar w} h_1 \p _w h_2\, , \qquad
W =  \p _w h_1 \p _{\bar w} h_2 + \p _{\bar w} h_1 \p _w h_2\, , 
\cr
&&
N_1  =  2 h_1 h_2 |\p_w h_1|^2 - h_1 ^2 W\, , \qquad
N_2  =  2 h_1 h_2 |\p_w h_2|^2 - h_2 ^2 W,
\eea
where $w$ is an arbitrary complex coordinate on $\Sigma$.
Then we have
\bea
&& e^{2\varphi}=-\frac{N_2}{N_1}\, , \qquad
\rho^8=-\frac{W^2 N_1 N_2}{h_1^4 h_2^4}\, , 
\cr && 
f_1^4=-4 e^\varphi h_1^4\frac{W}{N_1}\, , \qquad
f_2^4=4 e^{-\varphi} h_2^4\frac{W}{N_2}\, , \qquad
f_4^4=4 e^{-\varphi} \frac{N_2}{W}\, , 
\eea
while the relevant components of the two- and four-form 
fields (\ref{ansatz-forms}) are
\bea
&& b_1=-2i\frac{h_1^2 h_2 V}{N_1}-2\tilde h_2\, , 
\qquad b_2=-2i\frac{h_1 h_2^2 V}{N_2}+2\tilde h_1,
\label{forms-comp}
\eea
as given in \cite{D'Hoker:2007fq}, and
\ba
j_2=i h_1 h_2 \frac{V}{W}+3i({\cal C}-\bar{{\cal C}})-
\frac{3}{2}(\tilde h_1 h_2-h_1 \tilde h_2) \, ,
\label{j2-soln}
\ea
as we show in Appendix \ref{sec-5form}.
The holomorphic function $\cC$ is defined implicitly by
\ba
\p_w \cC= \cA \p_w \cB- \cB \p_w \cA\, . \label{def-C}
\ea
 
The behavior of various quantities near the real axis ($y=0$)
was studied in \cite{D'Hoker:2007fq}:

\begin{center}
\begin{tabular}{c|c|cccccccccc}
Intervals & Vanishing fiber 
&$h_1$ &$\p_y h_1$& $h_2$ & $V$ & $W$ & $N_1$ & $N_2$
\\
\hline
$[e_{2I}, e_{2I-1}]$
&
$S^2$ & $\Ocal(1)$ & $\Ocal(y)$ &  $\Ocal(y)$ & $\Ocal(1)$ & $\Ocal(y)$ & 
$\Ocal(y)$& $\Ocal(y)$
\\
others
&
$S^4$& $\Ocal(y)$ &$\Ocal(1)$ & $\Ocal(y)$ & $\Ocal (y)$ &
$\Ocal(1)$ & $\Ocal(y^4)$ & $\Ocal(y^4)$
\end{tabular}
\end{center}
It follows that $b_2=2\tilde h_1=-4 {\rm Im}\, \Acal$
on $[e_{2I},e_{2I-1}]$.

\section{An explicit expression  for the four-form}
\label{sec-5form}

The component $j_2$ of the RR four-form $C_{(4)}$ (\ref{ansatz-forms}) 
 is not given explicitly in \cite{D'Hoker:2007fq}, but can be obtained 
 along the lines of the similar computation in Section 9.9 of 
\cite{D'Hoker:2007xy}. 
We use the notation of these papers.
 
The derivative of $j_2$ admits an expression\footnote{
Using complex coordinates on $\Sigma$, the frames become $e^8=e^w+e^{\bar w}$, $e^9=-i(e^w-e^{\bar w})$, with $e^w=\rho\, dw$ and $e^{\bar w}=\rho\, d\bar w$.}
\bea
dj_2=-i f^4_4\left(\rho f_w dw-\rho f_{\bar w}d\bar w\right),
\eea
where from eqs. (5.24) and (6.1) of \cite{D'Hoker:2007fq} and 
from the relation $\rho p_w=\partial_w\phi$ ($\phi\equiv \varphi/2$) one has
\bea
2 \rho f_w=\partial_w \log \frac{\bar \beta}{\bar\alpha}+\left(\frac{\beta\bar \beta}{\alpha\bar \alpha}-\frac{\alpha\bar\alpha}{\beta\bar\beta}\right)\partial_w\phi.
\eea
Using that 
\bea
\alpha=\sqrt{\frac{\bar \kappa}{\rho}}\sqrt{\cosh(\phi+\bar\lambda)}\, , \qquad
\beta=i \sqrt{\frac{\bar \kappa}{\rho}}\sqrt{\sinh(\phi+\bar\lambda)}\, ,
\eea
it becomes
\bea
2\rho f_w=\frac{\partial_w \phi+\partial_w\lambda}{\sinh(2\phi+2\lambda)}-\frac{2\cosh(\lambda-\bar\lambda)}{|\sinh(2\phi+2\lambda)|}\partial_w\phi\, .
\eea
The warp factor is given by $f_4=\nu(\bar\alpha \beta+\bar\beta\alpha)$ (with $\nu=\pm 1$), so that
\bea
f_4^4=\frac{\kappa^2\bar\kappa^2}{\rho^4}\left(\sinh(2\phi+\lambda+\bar\lambda)-|\sinh(2\phi+2\lambda)|\right)^2\, .
\eea 
One can now change the variables from $\phi$ (real) and $\lambda$ (holomorphic) to the real variables $\mu$ and $\vartheta$ defined by
\bea
\lambda-\bar\lambda=i\mu\, , \qquad e^{2 i \vartheta}=\frac{\sinh(2\phi+2\lambda)}{\sinh(2\phi+2\bar\lambda)}\, ,
\eea
from which also follows
\bea
|\sinh(2\phi+2\lambda)|^2=\frac{(\sin2\mu)^2}{4\sin(\vartheta+\mu)\sin(\vartheta-\mu)}\, , \qquad e^{-i\vartheta}=\frac{|\sinh(2\phi+2\lambda)|}{\sinh(2\phi+2\lambda)}\, ,
\eea
and 
\bea
\partial_w\phi=-\frac{\sin 2\mu \, \partial_w\vartheta}{4\sin(\vartheta+\mu)\sin(\vartheta-\mu)}-\frac{i}{2}\partial_w\mu+\frac{\sin2\vartheta\, \partial_w\mu}{4\sin(\vartheta+\mu)\sin(\vartheta-\mu)}\, .
\eea
Using eq. (7.4) of \cite{D'Hoker:2007fq} one has 
\bea
2 \rho f_w f^4_4 &=&\frac{1}{2\hat\rho^4 \cos^2\mu}\Big[e^{-i\vartheta}\left(-\sin2\mu\, \partial_w\vartheta-ie^{2 i \vartheta}\, \partial_w\mu+i\cos2\mu\, \partial_w\mu\right)-\cr
&& \hskip 2cm -2\cos\mu\left(-\sin2\mu\, \partial_w\vartheta+ie^{-2 i \vartheta}\, \partial_w\mu-i\cos2\mu\, \partial_w\mu\right)\Big]\, .
\eea
In terms of
$
\psi=\frac{\sin\mu}{\hat\rho^2}e^{-i\vartheta/2},
$
the expression above becomes
\bea
2\rho f_w f^4_4&=&\frac{2i}{(\sin2\mu)^2}\Big[-\sin2\mu\left(\psi\, \partial_w\psi-\psi^2\, \partial_w\bar\psi\, /\bar\psi\right)-\bar\psi^2\, \partial_w\mu+\cos2\mu\, \psi^2\, \partial_w\mu+\cr && \hskip 2cm +2\cos\mu\sin2\mu\left(\bar\psi\, \partial_w\psi-\psi\, \partial_w\bar\psi\right)-2\cos\mu\, \psi^3\, \partial_w\mu\, /\bar\psi+\cr
&& \hskip 2cm +2\cos\mu\cos2\mu\, \psi\bar\psi\, \partial_w\mu\Big]\, ,
\eea
and finally, using the equation of motion 
\bea
\partial_w\bar\psi=\cot\mu\, \bar\psi\, \partial_w\mu+\frac{1}{\sin\mu}\psi\, \partial_w\mu
\eea
to eliminate the pieces with more than 2 $\psi$ and/or $\bar \psi$,
\bea
2\rho f_w f^4_4&=&2i\Big[-\frac{\psi\, \partial_w\psi}{\sin 2\mu}+\frac{\psi^2-\bar\psi^2}{(\sin 2\mu)^2}\partial_w\mu+\frac{2\cos2\mu}{(\sin2\mu)^2}\psi^2\, \partial_w\mu+\cr && \hskip .7cm +\frac{2\cos\mu}{\sin2\mu}(\bar\psi\, \partial_w\psi-\psi\, \partial_w\bar\psi)+\frac{2\cos\mu\, \cos2\mu}{(\sin2\mu)^2}\psi\bar\psi\, \partial_w\mu\Big]\, . 
\eea 
This can be almost written as a total derivative
\bea
2\rho f_w f^4_4=\partial_w\left(2i\frac{\psi\bar\psi}{\sin\mu}-i\frac{\psi^2+\bar\psi^2}{\sin2\mu}\right)-3i\frac{\psi^2\, \partial_w\mu}{\sin^2\mu}\, ,
\eea
using again the equation of motion for $\bar\psi$. Using the equation of motion for $\psi$, eq. (7.7) of \cite{D'Hoker:2007fq}, the expression for $\kappa$ in eqs. (7.8) and (7.13) and the last equation in (7.14), the last term in the formula above becomes 
\bea
-\frac{\psi^2\, \partial_w\mu}{\sin^2\mu}=\partial_w\left(\psi^2 \cot\mu+ih^2_1e^{-2\bar\lambda}-ih_2^2e^{2\bar\lambda}\right)+2i(h_1\partial_w h_2-h_2\partial_w h_1)\, .
\eea
Then
\bea
\rho f_w f^4_4=-i\partial_w(h_1 h_2 \tan\mu)-3(h_1\partial_w h_2-h_2\partial_w h_1)\, ,
\eea
and one has
\bea
j_2=- h_1 h_2\tan\mu+3i({\cal C}-\bar{{\cal C}})-\frac{3}{2}(\tilde h_1 h_2-h_1 \tilde h_2)\, .
\eea
Using (\ref{def-C}) together with 
the relations $\mu=-i(\lambda-\bar\lambda)$ and $e^{2\lambda}=\partial_w h_1/\partial_w h_2$, one can rewrite this as (\ref{j2-soln}).


\section{Asymptotic behavior}
\label{sec-asymptotic}

Let us study the asymptotic forms of physical fields
in the region $z\ra\infty$.
We use the $SU(N)$ identification (\ref{SU-ident})
of the matrix model and geometry data.

From the definition (\ref{omega}),
$\omega$ behaves in the asymptotic region of $\Sigma$ as
\ba
\omega(z)=\f{\lambda}z +\Ocal\left(\oo{z^3}\right).
\ea
The order ${\cal O}(z^{-2})$ term vanishes in the $SU(N)$ case.
Using the formulas in Appendix \ref{sec-geom},
we find the asymptotic forms of various fields:
\ba
e^{\varphi}\equiv e^{2\phi}=g_s+\Ocal(r^{-4}),
\hspace{27mm}
\label{asymp-dilaton}\\
f_1=\left(\f{\alpha'^{2}}{g_s\lambda}\right)^{1/4}r+\Ocal(1/r),\qquad 
f_2=\left(\f{\alpha'^{2}}{g_s\lambda}\right)^{1/4}r+\Ocal(1/r),\\
f_4=\left(\f{\alpha'^2 \lambda}{g_s}\right)^{1/4} |\sin\theta|+\Ocal(1/r^2),
\hspace{18mm}
\\
b_1=\Ocal(1/r),\qquad
b_2=\Ocal(1/r),
\hspace{20mm}
\\
j_2=-\f{\alpha'^2\lambda[12\theta-8\sin(2\theta)+\sin(4\theta)]}{32 g_s}
+\Ocal(1/r^2),
\\
b_4=\Ocal(1/r).
\hspace{40mm}
\ea
Here we introduced polar coordinates $z=r e^{i\theta}$
with $-\pi \leq \theta\leq 0$.
Note that the metric (\ref{warped-metric})
is written in the Einstein frame where
the AdS radius is $( \alpha'^2 \lambda/g_s)^{1/4}$
in our convention (\ref{asymp-dilaton}) for the dilaton.
The subleading terms depend on the representation $R$
and can be easily calculated in terms of the moments
of the eigenvalue distribution.


\bibliography{bubloops}  

\end{document}